\documentclass[useAMS,usenatbib,letterpaper]{mn2e}

\usepackage{epsfig,bm}
\usepackage{amsbsy,amssymb,amsmath}
\usepackage{hyperref}
\usepackage{graphicx}
\usepackage{natbib}
\usepackage{float}

\newcommand{\be}{\begin{equation}}
\newcommand{\ee}{\end{equation}}
\newcommand{\bea}{\begin{eqnarray}}
\newcommand{\eea}{\end{eqnarray}}
\newcommand{\bk}{{\bf k}}
\newcommand{\bl}{{\bf l}}

\newcommand{\Npe}{N_\perp} 
\newcommand{\Npa}{N_\parallel}

\newcommand{\kpe}{k_\perp}
\newcommand{\kpa}{k_\parallel}

\newcommand{\bell}{\mathbf{l}}
\newcommand{\bL}{\mathbf{L}}
\newcommand{\bx}{\mathbf{x}}
\newcommand{\phis}{\phi^\star}

\title[21\,cm lensing at $z \sim 2-3$]
{Weak lensing with 21\,cm intensity mapping at $z \sim 2-3$ }
\author[ Pourtsidou  \& Metcalf ]{ A. Pourtsidou \& R. Benton Metcalf  \\ 
 Dipartimento di Fisica e Astronomia, Universit\'{a} di Bologna, viale B. Pichat 6/2  , 40127, Bologna, Italy}
\begin{document}

\maketitle

\begin{abstract}
We study how 21 cm intensity mapping can be used to measure gravitational lensing over a wide range of 
redshift. This can extend weak lensing
measurements to higher redshifts than are accessible with conventional galaxy 
surveys. We construct a convergence estimator taking into account the discreteness
of galaxies and calculate the expected noise level as a function of
redshift and telescope parameters.  At $z \sim 2-3$ we find that a
telescope array with a collecting area $\sim 0.2 \, {\rm km}^2$ 
spread over a region with diameter $\sim 2 \, {\rm km}$ would be
sufficient to measure the convergence power spectrum to high
accuracy for multipoles between 10 and 1,000. 
We show that these measurements can be used to constrain interacting dark energy models.
\end{abstract}


\section{Introduction}

We now live in an era of precision cosmology. Almost all of the
information used to achieve this precision has come from redshifts
below $z\sim ~1.5$ or from the Cosmic Microwave Background (CMB) at
$z\sim 1000$. The vast regions between these redshifts have been probed
only sparsely.  
Given our ignorance of what is causing the apparent
acceleration of the Universe, it is important that we explore the
evolution of expansion and structure formation over the widest
possible range of redshift.  It is possible that dark energy, or a
modification to general relativity, came into play at higher redshift
than the standard cosmological constant model predicts \citep{Cope06,Clifton12}. 
Early dark energy models are an example of this.
In recent years, several 21~cm surveys have been proposed
to study the epoch of reionization (EoR) which could provide some
cosmological information at $z \sim 8-12$.  The gravitational lensing
of the CMB also provides some information on the intermediate
redshifts, but the signal-to-noise is low. The clustering of quasars
and Ly-$\alpha$ absorption lines in quasar spectra
can be measured at high redshift, but here bias and modeling uncertainties are serious
problems.  In this paper we address
the prospects for measuring gravitational lensing at redshifts after
reionization, but before those probed by galaxy surveys in the 
visible bands.

In \cite{Zahn:2005ap} and \cite{Metcalf:2009}  it was shown that if the EoR is at redshift 
$z \sim 8$ or later, a large radio telescope such as SKA (Square Kilometer Array) could measure the lensing convergence power spectrum and constrain the standard cosmological parameters. The authors extended the
Fourier-space quadratic estimator technique, which was first developed by 
\citep{Hu:2001tn} for CMB lensing  observations to three dimensional
observables, i.e. the $21$ cm intensity field $I(\theta,z)$.  These
studies did not consider 21~cm observations from redshifts after
reionization when the average HI density in the universe is much smaller.

It has also been proposed that lensing could be measured at lower 
redshifts by counting the fluctuations in the number density of
detected 21~cm objects on the sky as a measure of the magnification \citep{Zhang05,Zhang06,Zhang11}.
The signal-to-noise is greatly reduced in this case
because of the low number density of objects and the intrinsic
clustering of them.

Lensing surveys in the visible are limited in redshift by the number density of
detected galaxies with measurable ellipticities.  This is strongly
dependent on the depth of the survey, but any proposed survey that
will cover a significant fraction of the sky will be quite sparse  in
sources above $z \sim 1.5$.  Here we show that  21 cm observations
can be used to extend weak lensing measurements to higher redshifts 
than this, but still well below the redshift of reionization or the CMB.

21~cm intensity mapping is a technique that has been proposed for measuring the
distribution of HI gas before and during reionization (see
\cite{Furlanetto2006} for a review) and measuring the BAO at redshifts 
of order unity \citep{Chang:2008,Chang:2010,Seo:2009fq, Masui:2010, Ansari2012,Battye:2012tg,Chen2012,Pober2013}. 
In this technique, no attempt is made
to detect individual objects.  Instead the 21~cm emission is treated
as a continuous three dimensional field.  The angular resolution of the
telescope need not be high enough to resolve individual galaxies which
makes observations at high redshift possible with a reasonably sized
telescope.  Foregrounds are expected to have smoother spectra than the
signal so they can be subtracted by filtering in frequency.

In this study, we extend the 21~cm lensing method further, taking into
account the discreteness of galaxies. In Section~\ref{sec:formalism},
we present our formalism for constructing a lensing estimator and
calculating the corresponding lensing reconstruction noise. In
Section~\ref{sec:results}, we investigate the possibility of measuring
lensing at intermediate redshifts and show results using telescope
arrays optimized for high signal-to-noise. Measurements of the convergence power spectrum
can be used to constrain interacting dark energy models. We
conclude in Section~\ref{sec:concl}.

\section{Formalism}
\label{sec:formalism}

The mean observed brightness temperature at redshift $z$ due to the average HI density can be written as
\citep{Battye:2012tg}
\be
\label{eq:Tz}
\bar{T}(z) = 180 \; \Omega_{\rm HI}(z) \; h \frac{(1+z)^2}{E(z)} \; {\rm mK},
\ee where the Hubble parameter $h=H_0/100\,{\rm km}\,{\rm s}^{-1}\,{\rm Mpc}^{-1}$, $E(z)=H(z)/H_0$ and
$\Omega_{\rm HI}(z)=8\pi G \rho_{\rm HI}(z)/(3H^2_0)$ is the average HI density at redshift $z$ relative to the present day critical density. 
Consequently, the 3D HI power spectrum of the brightness temperature fluctuations is given by
\be
P_{\Delta T_b}(k) = [\bar{T}(z)]^2 (1+f\mu^2_k)^2 P_\delta(k),
\ee where $P_\delta(k)$ is the underlying dark matter power spectrum,
$f = \frac{d \ln D}{d\ln a} \simeq \Omega_m(z)^{0.55}$  where $D$ is the
linear growth rate and $\mu_k$ is the cosine of the angle between the
wave vector $\bk$ and the line of sight $\hat{z}$.  The scale
parameter is $a = (1+z)^{-1}$.

In \cite{Zahn:2005ap} and \cite{Metcalf:2009} the  convergence
estimator and the corresponding lensing reconstruction noise are
calculated assuming that the temperature (brightness) distribution is
Gaussian. The advantage of 21cm lensing is that one is able to combine
information from multiple redshift slices. In Fourier space, the
temperature fluctuations are divided into perpendicular to the line of
sight wave vectors $\mathbf{\kpe}=\mathbf{l}/{\cal D}$, with ${\cal
  D}$ the angular diameter distance to the source redshift, and a
discretized version of the parallel wave vector $\kpa =
\frac{2\pi}{{\cal L}}j$ where ${\cal L}$ is the depth of the observed
volume. Considering modes with different $j$ independent, an optimal
estimator can be found by combining the individual estimators for
different $j$ modes without mixing them. The three-dimensional lensing reconstruction noise is then found to be \citep{Zahn:2005ap}
\be
N(\bL) =  \left[\sum_j^{j_{\rm max}} \frac{1}{L^2}\int \frac{d^2\ell}{(2\pi)^2}  \frac{[\bell \cdot \bL C_{\ell,j}+\bL \cdot (\bL-\bl)
C_{|\ell-L|,j}]^2}{2 C^{\rm tot}_{\ell,j}C^{\rm tot}_{|\bl-\bL|,j}}\right]^{-1},
\ee where
\be
\label{eq:Cellj}
C_{\ell,j} = \frac{P_{\Delta T_b}(\sqrt{(\ell/{\cal D})^2+(j2\pi/{\cal L})^2})}{{\cal D}^2 {\cal L}} =  [\bar{T}(z)]^2  P_{\ell,j}.
\ee

However, the Gaussian case is an approximation which breaks down if we
take into account the discreteness of galaxies in the Universe. After
reionization, the HI resides mostly in the galaxies. A more realistic
model the HI distribution, and the one most often assumed, is a Poisson distribution drawn from a Gaussian distribution representing
the clustering of galaxies.
In order to calculate a lensing estimator and the corresponding lensing reconstruction noise for this model, we will work with the discrete Fourier transform of the intensity field $I(\bx)$, which we write as
\be
I_{\bk} = \frac{\Omega_s}{\Npe \Npa} \sum_{\bx}e^{i\bk \cdot \bx} I(\bx),
\ee where $\bk=(\bell,j)$, $\bx=(\theta,z)$ and $\Omega_s = \Theta_s \times \Theta_s$ for a square
survey geometry.   $\Npe$ and $\Npa$ are the number of cells in the
volume perpendicular and paralel to the radial direction.  We also have
\be
I(\bx) = \frac{1}{\Omega_s}\sum_{\bk}e^{-i\bk \cdot \bx}I_{\bk}.
\ee
For the $2$-point correlation function we get
\be
<I(\bx) I(\bx') > = \frac{1}{\bar{\eta}\delta V}\frac{<M^2>}{<M>^2}\delta^K_{\bx\bx'} + \xi_{\bx \bx'}.
\ee
Fourier transforming we find
\be
<I_{\bk} I^*_{\bk'}> = \Omega_s \; (P_{\ell,j}+P^{\rm shot}) \; \delta^{K}_{\ell,\ell'}\delta^{K}_{jj'},
\ee
where $P_{\ell,j}$ is given by Equation~(\ref{eq:Cellj}) and 
\be
P^{\rm shot} = \frac{1}{\bar{\eta}}\frac{1}{D^2{\cal L}}\frac{<M^2>}{<M>^2},
\ee with $\bar{\eta}$ the average number density of galaxies and the $M$ moments must be computed
from an appropriate mass (or luminosity) function. 
The lensing correlation gives
\begin{align} \nonumber
&<\tilde{I}_{\bell,j}\tilde{I}^*_{\bell-\bL,j'}> = \delta^K_{jj'} \times  \\
&[\bell \cdot \bL P_{\ell,j}+\bL \cdot (\bL-\bl)P_{|\ell-L|,j} 
+ L^2 \; P^{\rm shot} ]\; \Psi(\bL).
\end{align}
We can construct a lensing estimator of the form
\be
\hat{\Psi}(\bL)= f(\bL) \sum_j^{j_{\rm max}} \sum_{\bell} \tilde{I}_{\bell,j}\tilde{I}^*_{\bell-\bL,j},
\ee where $f(\bL)$ is a normalization. In order for the estimator to be unbiased we impose
\be
<\hat{\Psi}(\bL)> = \Psi(\bL),
\ee and we find (note $P_{\ell,j} \rightarrow C_{\ell,j}$ from now on, as in Eq.~(\ref{eq:Cellj}))
\begin{align}
f(\bL)=\bigg\{\sum_j^{j_{\rm max}} \sum_{\bell}[\bell \cdot \bL C_{\ell,j}+\bL \cdot (\bL-\bl)
C_{|\ell-L|,j}+L^2 \; C^{\rm shot}]\bigg\}^{-1},
\end{align} with $C^{\rm shot} =  [\bar{T}(z)]^2 P^{\rm shot}$.

We are now ready to compute the lensing reconstruction noise $N(L)$, which corresponds to the
variance of the estimator ${\cal V}= <\hat{\Psi}(\bL) \hat{\Psi}^{\star}(\bL)>$.
After some algebra and using
\be \nonumber
\sum_{\bell} \rightarrow \Omega_s \int \frac{d^2\ell}{(2\pi)^2}
\ee to move from discrete to continuous $\ell$-space we find
\begin{align} \nonumber
& N(L) = L^2  \times \\
& \frac{{\cal N}_0+{\cal N}_1+{\cal N}_2+{\cal N}_3+{\cal N}_4}
{\bigg\{\displaystyle\sum_j^{j_{\rm max}} \int \frac{d^2\ell}{(2\pi)^2}[\bell \cdot \bL C_{\ell,j}+\bL \cdot (\bL-\bl)
C_{|\ell-L|,j}+L^2 \; C^{\rm shot}]\bigg\}^2} \, ,
\label{eq:NL}
\end{align} with
\be \nonumber
{\cal N}_0=
[\bar{T}(z)]^4 (j_{\rm max})^2 \frac{1}{\bar{\eta}^3}\frac{1}{(D^2{\cal L})^3}\frac{<M^4>}{<M>^4}\left(\int\frac{d^2\ell}{(2\pi)^2}\right)^2,
\ee
\begin{align} \nonumber
{\cal N}_1= [\bar{T}(z)]^2 & (j_{\rm max}) \frac{1}{\bar{\eta}^2}\frac{1}{(D^2{\cal L})^2}\frac{<M^3>}{<M>^3} \left(\int\frac{d^2\ell'}{(2\pi)^2}\right) \\ \nonumber
&\times \sum_j^{j_{\rm max}} \int\frac{d^2\ell}{(2\pi)^2} [2C^{\rm tot}_{\ell,j}+2C^{\rm tot}_{|\ell-L|,j}],
\end{align}
\begin{align}  \nonumber
{\cal N}_2 = [\bar{T}(z)]^2 &
\frac{1}{\bar{\eta}^2}\frac{1}{(D^2{\cal L})^2}\frac{<M^2>^2}{<M>^4}  \sum_j^{j_{\rm max}} \sum_{j'}^{j_{\rm max}} \int\frac{d^2\ell}{(2\pi)^2} \int\frac{d^2\ell'}{(2\pi)^2}  \\ \nonumber
& \;\;\;\;\;\;\;\;\;\;\;\;\;\;\;\;\;\;\;\;\;\; \times [C^{\rm tot}_{|\ell-\ell'|,|j-j'|}+C^{\rm tot}_{|\ell+\ell'-L|,j+j'}],
\end{align}
\begin{align} \nonumber
{\cal N}_3 &= C^{\rm shot} \sum_j^{j_{\rm max}} \int\frac{d^2\ell}{(2\pi)^2} [2C^{\rm tot}_{\ell,j}+2C^{\rm tot}_{|\ell-L|,j}]
\end{align} and
\begin{align} \nonumber
{\cal N}_4 &= \int\frac{d^2\ell}{(2\pi)^2} 2C^{\rm tot}_{\ell,j}C^{\rm tot}_{|\ell-L|,j} \, ,
\end{align}
where $C^{\rm tot}_{\ell,j} = C_{\ell,j}+C^{\rm N}_\ell$, with $C^{\rm N}_\ell$ the thermal noise of the telescope.

In the next section, we will use the constructed estimator and noise
to investigate how well the convergence power spectrum can be measured
from data as a function of telescope parameters. Note that the derived noise contains the HI mass moments (up to 4th order), which need to be calculated assuming an adequate mass function. The most interesting feature of Eq.~(\ref{eq:NL}) is that the shot noise terms contribute to both the noise and the signal in the lensing measurement. 

A significant difficulty in 21 cm experiments is foreground
contamination from galactic synchrotron, point sources, bremsstrahlung
etc. These foreground contributions are smooth power laws in frequency, and it is
expected that they can be removed to high accuracy.  We will present a study of foreground removal in a
future paper. For now, we note that foreground removal will make the first few $j$-modes useless for the reconstruction \citep{Zahn:2005ap}, so we have discarded the $j = 0$ mode in our calculations.

\section{Results}
\label{sec:results}

In general, there are three main epochs of interest: (i) the Dark Ages before reionization, where the HI fraction is high but so are the foregrounds and noise (ii) the EoR (iii) the epoch after reionization. During the latter epoch, the HI fraction is much lower ($\sim 1\%$ today), but the foregrounds and noise are also lower.

For this work we will concentrate on the last epoch and work at a redshift $z_s=2$, which corresponds to a frequency $\nu=473 \, \rm{MHz}$ (detailed work on all three epochs will be presented in a future paper). 
Considering a uniform distribution ground based array
of telescopes, the power spectrum of the thermal noise will be
\be
C^{\rm N}_\ell = \frac{(2\pi)^3 T^2_{\rm sys}}{B t_{\rm obs} f^2_{\rm cover} \ell_{\rm max}(\nu)^2} \, ,
\ee where 
$T_{\rm sys}$ is the system temperature, $B$ is the bandwidth, $t_{\rm obs}$ the total observation time, $D_{\rm tel}$ the diameter of the array and $\ell_{\rm max}(\lambda)=2\pi D_{\rm tel}/\lambda$ is the highest multipole that can be measured by the array at frequency $\nu$ (wavelength $\lambda$). $f_{\rm cover}$ is the total collecting area of the telescopes $A_{\rm coll}$ divided by $\pi(D_{\rm tel}/2)^2$, the aperture covering fraction. Our chosen telescope configuration follows a SKA-like design. The total collecting area is $\sim 0.19 \; {\rm km}^2$ (30\% of the full SKA) and the maximum baseline is $D_{\rm tel}=2 \; {\rm km}$, giving an 
$f_{\rm cover} \simeq 0.06$ and a value of $\ell_{\rm max} \sim
19900$.  We consider a $2 \, {\rm yr}$
observation time and a $40 \, {\rm MHz}$ bandwidth. 
Note that the change of the convergence power spectrum across the
corresponding redshift interval is very small. This would not be the case at a much higher redshift (e.g. $z\sim 8$), where we would have to use smaller bandwidths $\sim 1 \, {\rm MHz}$.

The most important source of noise is Galactic synchrotron emission, approximated by
\be
T_{\rm syn} = 180 \; {\rm K} \; (\nu/180 \; {\rm MHz})^{-2.6}.
\ee
However, at $z=2$ this is subdominant in comparison to the receiver temperatures which
we estimate to be $\sim 50 \, {\rm K}$, and this is the value we are going to use for $T_{\rm sys}$.

In order to calculate the Poisson terms we need the HI mass function. The comoving number density of galaxies $dn$ in a mass range $dM$ is taken to be a Schechter function
\be
\frac{dn}{dM}dM = \phis \left(\frac{M}{M^\star}\right)^\alpha {\rm exp}\left[-\frac{M}{M^\star}\right]\frac{dM}{M^\star},
\ee
parametrized by a low-mass slope $\alpha$, a characteristic mass $M^\star$ and a normalization $\phis$. We can calculate $\rho_{\rm HI}$ using
\begin{align} \nonumber
\rho_{\rm HI} &=  \phis M^\star \int \left(\frac{M}{M^\star}\right)^{\alpha+1} {\rm exp}\left[-\frac{M}{M^\star}\right]\frac{dM}{M^\star} \\
& = \phis M^\star \;\Gamma(\alpha+2),
\end{align} where $\Gamma$ denotes the Gamma function. 
The HI mass density relative to the critical density of the Universe $\rho_{\rm c} = 2.7755 \, h^2 10^{11}  \, M_\odot \, {\rm Mpc}^{-3}$ is
\be
\Omega_{\rm HI} = \frac{\rho_{\rm HI}}{\rho_{\rm c}}= \frac{\phis M^\star \;\Gamma(\alpha+2)}{\rho_{\rm c}},
\ee and is used in Equation~(\ref{eq:Tz}) to calculate $\bar{T}(z)$.

The parameters  ($\alpha, M^\star, \phis$) are the most important source of systematic uncertainty in
our study. They are only well measured in the local Universe.  We assume a no-evolution model using the values
$\alpha=-1.3, M^\star=3.47h^{-2}10^9 \, M_\odot, \phis = 0.0204 \, h^3
\, {\rm Mpc}^{-3}$ reported from the HIPASS survey \citep{Zwaan03}.
Other models derived from  Lyman-$\alpha$ systems are possible (see, for example, \citep{Peroux:2001ca}),
but we feel that no-evolution is a conservative choice.

One of the first objectives of a $21\rm{ cm}$ lensing survey will be to measure the two-point statistics of the convergence field $\kappa(\vec{L},z_s)$ or, equivalently, the displacement field $\delta \theta(\vec{L},z_s)$, averaged over $z_s$. That is,
\be
C_L^{\delta \theta \delta \theta} = \frac{9 \Omega^2_m H^3_0}{L(L+1) c^3} \int_0^{z_s}dz \, P_{\delta}(k=L/{\cal D}(z),z) [W(z)]^2 /E(z),
\ee where $W(z)=({\cal D}(z_s)-{\cal D}(z))/{\cal D}(z_s)$.
The expected error
in the power spectrum $C_L^{\delta \theta \delta \theta}$ averaging over $\bL$ directions in a band of width $\Delta L$ is given by \be
\label{eq:DCL}
\Delta C_L^{\delta \theta \delta \theta} =\sqrt{\frac{2}{(2L+1)\Delta L f_{\rm{sky}}}}\left(C_L^{\delta \theta \delta \theta}+N_L\right).
\ee 
There is a limit to the number of $j$-modes that can be used in
$N_L$.  For very high $j$ the internal velocity structure of galaxies
will be resolved and our statistical model which treats them as point
sources will break down.   To find the maximum $j$, we use the formula
$\Delta v/c = B/f$ to calculate the velocity width corresponding to our chosen bandwidth $B$ at the observed frequency $f$, and then we divide with a typical velocity dispersion for a galaxy at $z=2$ ($f=473 \, {\rm MHz}$), which we assume to be $200 \, {\rm km/s}$. 
This gives $j_{\rm max}=126$, but the noise has already
converged at $j \sim 40$. 

\begin{figure}
\centerline{
\includegraphics[scale=0.35]{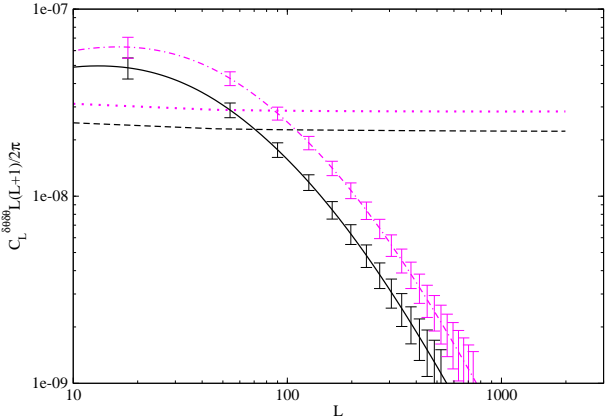}}
\caption{Displacement field power spectrum (solid black line) and
  lensing reconstruction noise $N_L$ (dashed black line, equation~(\ref{eq:NL})) for the
  compact SKA-like telescope described in the text, at redshift
  $z=2$.  $N_L$ converges well before $j$ reaches a reasonable $j_{\rm max}$, making the results insensitive to the
  exact value of this parameter. 
  The measurement errors come from
  sample variance and $N_L$ according to Eq.~(\ref{eq:DCL}). 
  We have
  chosen  $f_{\rm sky}=0.2$ and $\Delta L=36$. Note that the signal
  can be probed up to a much lower value than $\ell_{\rm max} \sim
  19900$, the highest multipole the telescope can
  reach.  We also show the results for redshift $z=3$ (dot-dashed and dotted magenta lines).}
\label{CL}
\end{figure}

In Fig.~\ref{CL} we compare the signal (solid black line), i.e. the
displacement field power spectrum $C_L^{\delta \theta \delta \theta}$,
with the noise $N_L$ (dashed black line).  As in the Gaussian case,
the shape of $L^2 N_L$ approaches a constant --- it does diverge in very high multipoles due to the thermal noise.
The measured lensing power spectrum will also depend on the multipole binning $\Delta L$ and the fraction of the sky surveyed $f_{\rm sky}$, as shown from Eq.~(\ref{eq:DCL}). Choosing $f_{\rm sky}=0.2$ and $\Delta L=36$ we get the 
measurement errors shown in Fig.~\ref{CL}. Repeating the calculation assuming the sources are at redshift $z_s=3$, we get the
results shown in Fig.~\ref{CL} for the signal $C_L^{\delta \theta \delta \theta}$ (dot-dashed magenta line) and the noise $N_L$ (dotted magenta line).

\begin{figure}
\centerline{
\includegraphics[scale=0.55]{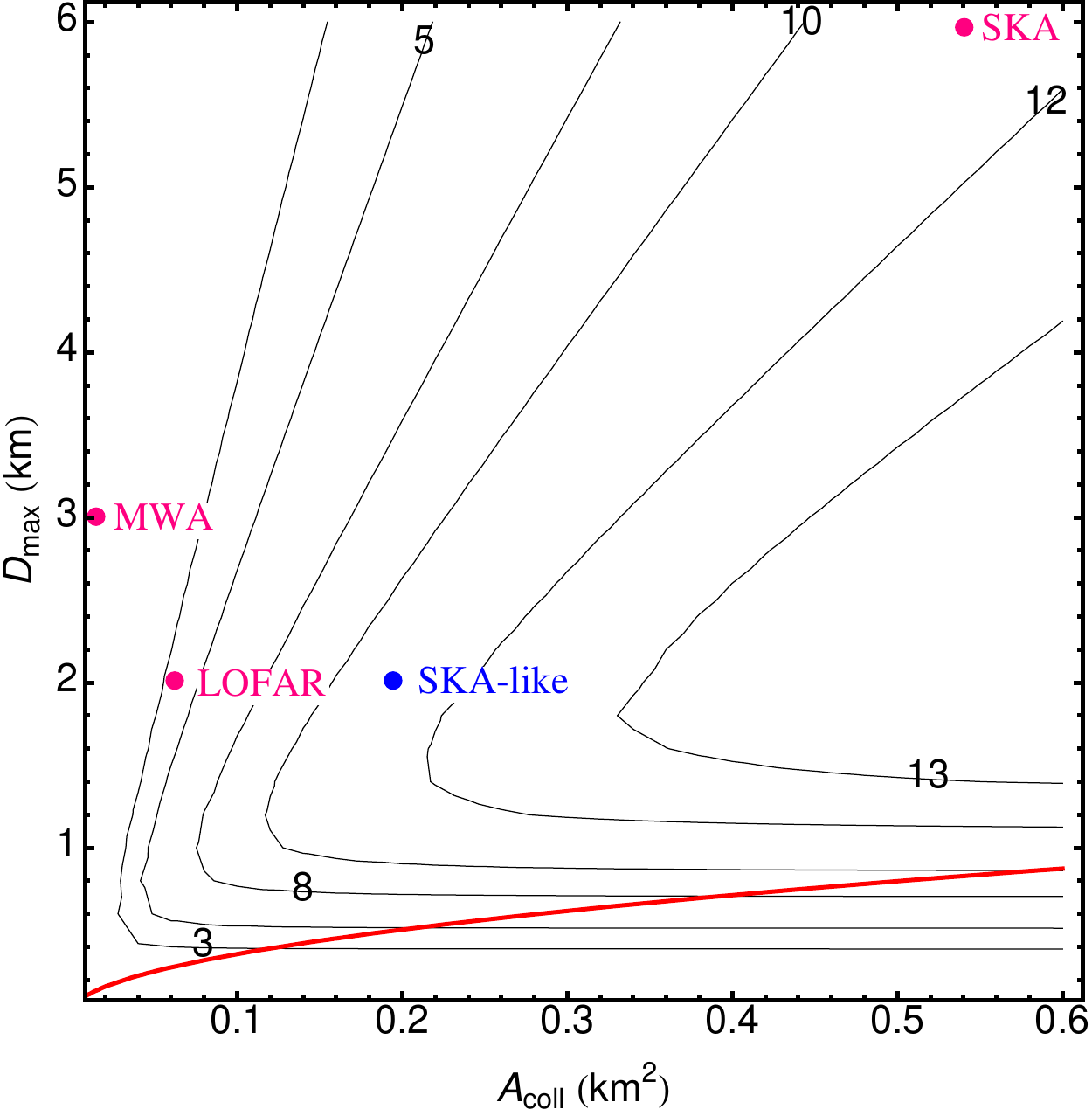}}
\caption{The signal-to-noise  (\ref{eq:DCL}) at $L=100$ for various
  telescope configurations. Sources are at $z=2$. The contour lines are labelled with the
  $(S/N)$ values. The area under the red solid line is excluded, since
  it corresponds to $f_{\rm cover} > 1$.  Some telescopes are shown
  for comparison although MWA and LOFAR do not operate at the required
frequency for this experiment. }
\label{contour}
\end{figure}

In Fig.~\ref{contour} we show the signal-to-noise (S/N) values at
multipole $L=100$ spanning the parameter space $(D_{\rm max},A_{\rm
  coll})$.  A LOFAR-like telescope could in principle
give good results, but it does not operate at  the right frequencies
to observe at $z=2$. The sparse SKA core array with $D_{\rm max}= 6 \,
{\rm km}$ gives a high S/N value at $L=100$, but the more compact
SKA-like configuration we have chosen performs better when one
computes the \emph{total} noise (i.e. taking into account the
contributions at all $L$), due to its higher covering fraction.

\begin{figure}
\centerline{
\includegraphics[scale=0.35]{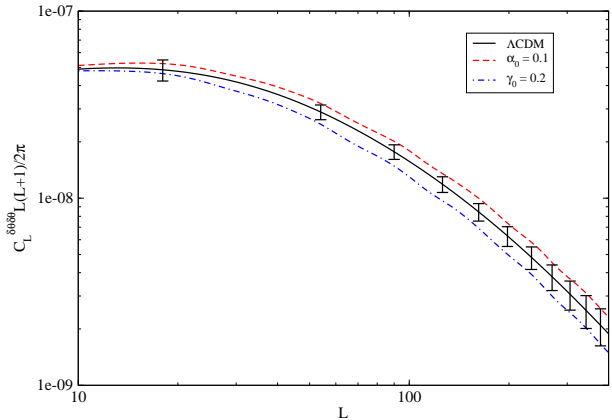}}
\caption{Displacement field power spectrum for $\Lambda$CDM compared
  with two different interacting dark energy models.  Sources are at $z=2$.  The error bars
  are the same as in figure~\ref{CL}.}
\label{coupled}
\end{figure}

Measurements of the weak lensing signal, such as those presented in Fig.~\ref{CL}, can be used to constrain interactions in the dark 
sector.  To illustrate this point we will adopt several concrete dark
energy models. 
 \cite{Pourtsidou:2013} found three distinct classes of dark
 energy models in the form of a scalar field $\phi$ coupled to cold
 dark matter (subscript cdm). 
The first two types involve energy and momentum transfer between the
dark sectors, while the third is a pure momentum transfer model.  The
coupled quintessence (CQ) model suggested by \cite{Amendola2000}
belongs to the Type-1 class.  In such a model, the Bianchi identities can be written as
\be
\nabla_\nu T^\nu_{(\phi) \mu} = - J_\mu = - \nabla_\nu T^\nu_{({\rm cdm}) \mu},
\ee so that the total energy-momentum tensor of the dark sector is conserved. The CQ Type-1 model has a coupling current
\be
J_\mu = - \alpha_0 \rho_{\rm cdm} \nabla_\mu \phi,
\ee 
where $\alpha_0$ is a constant
coupling parameter and $\rho_{\rm cdm} = \rho_{\rm cdm,0} a^{-3} e^{\alpha_0 \phi}$ is the CDM density for this model. We also consider a single exponential potential $V(\phi)$ for the 
quintessence field.
 Using a modified version of the \texttt{CAMB}
code \citep{camb} we can study the background cosmology and the linear
perturbations of the chosen model \cite[for details,
see][]{Pourtsidou:2013}.  We construct the
displacement field power spectrum and compare it with the $\Lambda$CDM
prediction in Fig.~\ref{CL}. Note that each cosmology evolves to the
PLANCK cosmological parameter values \citep{Ade:2013}. As we can see in
Fig.~\ref{coupled}, the Type-1 model with a coupling parameter
$\alpha_0=0.1$ would be excluded. Here there is energy transfer from
dark matter to dark energy making the dark matter density larger in
the past compared to the non-interacting case for fixed $\Omega_m$ today, hence the
gravitational potential is higher and the convergence power spectrum
is enhanced. The Type-3 class of models in \citep{Pourtsidou:2013} is
particularly interesting, as the background energy densities evolve as
in the uncoupled case. More specifically, in Type-3 models no coupling
appears in the fluid equations at the background level.
Furthermore, the energy-conservation equation remains uncoupled also at the linear level, so we have a pure momentum-transfer coupling at the level of linear perturbations. Working with the CQ Type-3 case studied in \citep{Pourtsidou:2013}, we find that the lensing signal is suppressed and a model with coupling parameter $\gamma_0=0.2$ would be excluded (see Fig.~\ref{coupled}).

\section{Discussion and conclusions}
\label{sec:concl}

Past work has been more pessimistic on the prospects of measuring
lensing from 21~cm radiation at the redshifts discussed here 
\citep{Zhang05,Zhang06,Zhang11}.  We believe that this is because
those studies were based on counting the number of galaxies
that are several sigma above the noise.  With that approach the
clustering of galaxies and the shot noise from their discreteness
contribute purely to noise in the lensing estimator.  In our approach,
shot noise and clustering contribute to both the noise and to an
improvement in the signal.  Surprisingly, lensing can be measured
without resolving (in angular resolution not frequency) or even
identifying individual sources.

We have developed a technique for measuring gravitational lensing in
21~cm observations of HI after reionization that takes into account the
discreteness of galaxies and find that it is very promising as a
method for measuring the evolution of the matter power spectrum at high
redshift. We have shown results here for two redshifts, but the
technique is applicable to any redshift below reionization with
varying degrees of signal-to-noise and could be used for tomographic
lensing studies by combining redshifts. In future work, we will develop this concept further by extending our
calculations to different redshifts, telescope configurations, models
for the high redshift HI mass function and foreground subtraction.

 \section*{Acknowledgments}

This research is supported by the project GLENCO, funded under the
FP7, Ideas, Grant Agreement n. 259349. 


\bibliographystyle{natbib}

\end{document}